%% file: main.tex
\input{header}

\begin{document}

\newcommand{\LG}[1]{\textcolor{red}{[#1]}}

\title{An Event-Driven Cloud-Native Wearable Analytics Framework for Real-Time Clinical Workloads}

\author{
\IEEEauthorblockN{Elias Grünewald$^1$*, Daniil Cherepko$^2$, Linus Gustafsson$^2$, Jakob Möhler$^2$, Oskar Rabe$^2$,\\ Paul Robin Reichelt$^2$, Constantin Stahl$^2$, Lukasz Sztukiewicz$^2$, Louis Agha-Mir-Salim$^1$, and Felix Balzer$^1$}
\IEEEauthorblockA{$^1$Institute of Medical Informatics, Charité -- Universitätsmedizin Berlin, Germany\\
$^2$Technische Universität Berlin, Germany\\
* elias.gruenewald@charite.de}
}

\maketitle
\pagestyle{plain}

\pagestyle{plain}

\begin{abstract}
Continuous physiological monitoring using consumer-grade wearables offers a transformative opportunity for clinical care and research, yet integration remains hindered by device heterogeneity, proprietary data formats, and strict regulatory requirements. We present an event-driven, cloud-native system designed to ingest, normalize, and analyze high-frequency vital signs from wearables at scale and without vendor lock-in. The system design proposes a multi-layered microservice architecture using cluster orchestration. Data acquisition is handled via a cross-platform mobile application that leverages native health frameworks, ensuring compatibility across fragmented device ecosystems. To address interoperability, we implement an event-driven transformation pipeline using stream processing engines and specialized services to map raw measurements to the FHIR standard for medical interoperability. Our novel dependency-aware FHIR minimization scheme reduces storage overhead while maintaining lossless resource reconstruction. Furthermore, the platform integrates a modular data analytics and machine learning layer based on a medallion lakehouse architecture, supporting the full machine learning lifecycle from real-time stream processing to model serving. Performance evaluation demonstrates that the ingestion pipeline sustains 50 full ingestion requests per second with median response times under 8 ms, satisfying the low-latency requirements for real-time patient monitoring. Our open-source implementation adheres to regulatory compliance standards through role-based access control and secure service-to-service communication, providing a robust foundation for deploying wearable-based monitoring in institutional healthcare settings for clinical decision support and research workloads.
\end{abstract}

\begin{IEEEkeywords}
Wearables, Cloud Native, Stream processing, FHIR, Clinical Decision Support
\end{IEEEkeywords}

\section{Introduction}
\noindent Modern healthcare increasingly relies on continuous physiological monitoring to detect complications early and improve patient safety and outcomes.
Wearable devices, such as smartwatches and smartbands, can record a broad spectrum of clinically relevant signals (e.g., heart rate, blood oxygen saturation, respiratory rate, and activity metrics) continuously and non-invasively, integrating seamlessly into patients' daily lives. This way, they have the potential to extend monitoring beyond traditional hospital settings, enabling early detection of complications before they occur (prevention) or during recovery after hospitalization \cite{ferguson2022effectiveness}. However, the integration of wearable data into clinical workflows and research has been hindered by several challenges.

Within healthcare institutions, clinical data flows through a complex landscape of patient data management systems, hospital information systems, and public health registries. Within hospitals, medical-grade devices are deployed and integrated into clinical workflows, particularly in critical care settings. Meanwhile, data that can be collected from wearable devices are often fragmented and lacks standardization. Such devices are often used in an ad-hoc manner, and data are not systematically collected or integrated into clinical system infrastructures \cite{doukas2011managing}. Routine data in hospitals already suffer from high technical barriers to interoperability, labor-intensive data extraction, uncertain data quality, inconsistent preprocessing, and limited generalizability \cite{Boie2026}. Consumer wearables from multiple manufacturers are each using proprietary formats and APIs. Adding such a heterogeneous data source further exacerbates the mentioned problems. Hence, raw wearable data must be normalized to a common interchange standard compatible with existing hospital information systems, access must be restricted to authorized clinicians and researchers, and the infrastructure must satisfy privacy, security, and medical device compliance requirements expected of systems handling sensitive health information \cite{hao2017role}.

In addition, there is a currently unmet need to support a wide range of use cases, from real-time clinical decision support to retrospective research analysis. Real-time applications require low-latency data processing and inference capabilities, while research workloads may involve large-scale data storage and batch processing \cite{jamiesonGuideConsumergradeWearables2025}. Additionally, the system must be flexible enough to accommodate evolving data sources, analytical methods, and clinical requirements over time.

To date, there is a lack of comprehensive frameworks that address these challenges in an integrated manner. Existing solutions often focus on specific aspects, such as data ingestion or analytics, without providing a holistic approach to managing the entire lifecycle of wearable data in healthcare settings. Wearable device vendors typically offer proprietary platforms that may not align with clinical standards or integrate seamlessly with hospital systems, further complicating adoption.

\begin{figure}[ht]
    \centering
    \includegraphics[width=1\linewidth]{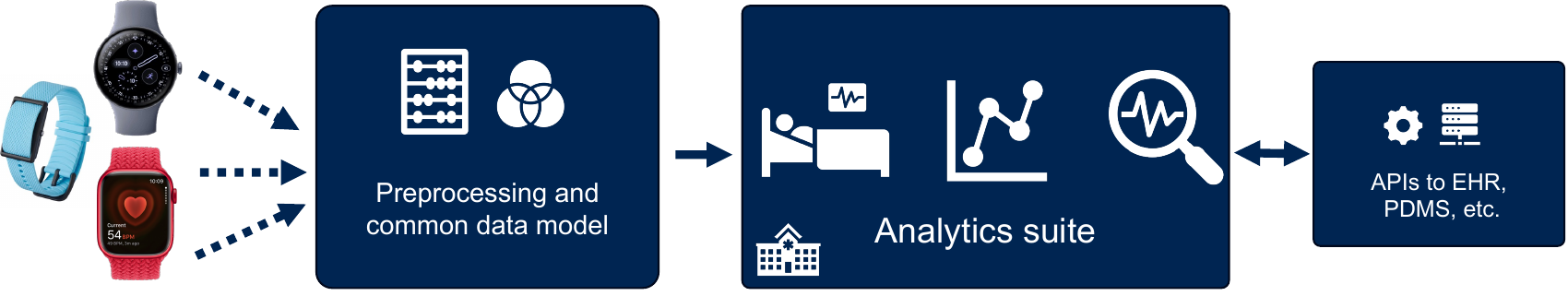}
    \caption{Conceptual view of proposed wearable data integration.}
    \label{fig:high-level}
\end{figure}

This situation motivates two research questions: How can we design a scalable and flexible architecture that supports the ingestion, processing, and analysis of heterogeneous wearable data in real-time? What strategies can be employed to ensure interoperability with existing healthcare systems while maintaining privacy and compliance?

To address these multifaceted challenges, we propose a cloud-native framework that leverages modern cloud and data engineering principles to facilitate the ingestion, processing, and analysis of wearable data in real-time. Decoupling ingestion, interoperability, and AI-inference workloads through a microservices architecture allows for flexible scaling and maintenance. By mapping heterogeneous wearable data payloads to clinically relevant schemata, we ensure interoperability with existing healthcare systems and enable seamless integration into clinical workflows. In this paper, we contribute:
\begin{itemize}
    \item A scalable, event-driven architecture for wearable data analytics that supports real-time inference and machine learning workloads, and
    \item A mapping of heterogeneous wearable data payloads to clinically meaningful schemata, enabling interoperability with existing healthcare through standardized and customizable interfaces, and
    \item A fully functional and open-source prototype implementation of the proposed framework, and
    \item A demonstration of the framework's capabilities through benchmarking experiments. %
\end{itemize}

The remainder of this paper is organized as follows: Sect.~\ref{sec:related-work} reviews relevant background and related work in cloud-native and event-driven architectures, wearable data processing, and healthcare interoperability. Sect.~\ref{sec:system-design} describes the high-level architecture and design principles of our proposed framework. Sect.~\ref{sec:implementation} details the implementation of each layer of the system. Sect.~\ref{sec:evaluation} presents the evaluation methods and results, including a clinical case study. Finally, Sect.~\ref{sec:discussion} discusses the implications of our findings, limitations, future research directions, and concludes.

\section{Background \& Related Work}
\label{sec:related-work}

\noindent In this section, we review relevant background and related work in the areas of cloud-native architectures in clinical settings, healthcare interoperability, and wearable data processing.

\subsection{Cloud-Native Architectures in Health}
Healthcare systems are increasingly adopting cloud-native architectures to leverage the scalability, flexibility, and resilience offered by cloud computing \cite{zao2024design}. Cloud-native design principles, such as microservices, containerization, and orchestration, enable healthcare applications to handle variable workloads in clinical settings and integrate with diverse data sources \cite{avireneni_cloud-native_2025}. However, the adoption of cloud-native architectures in healthcare is still in its early stages and faces challenges in integration with legacy monolithic systems, in particular, tightly integrated hospital information systems \cite{ogeawuchi2023designing, dinh2019wearable}. Existing cloud-native architectures in healthcare often focus on specific use cases, such as electronic health record (EHR) management or telemedicine platforms, all too often without providing interoperability with patient data management systems, and the lack of a comprehensive framework for wearable data processing and analytics \cite{jamiesonGuideConsumergradeWearables2025}. To date, the potential of cloud-native architectures to support real-time clinical workloads and machine learning applications for immediate inference remains underexplored \cite{mulpuri2020ai}. Striving for data sovereignty and vendor independence, institutions begin to adopt cloud-native architectures on their own infrastructure, such as on-premises clusters or even private clouds, to maintain control over sensitive health data while benefiting from modern architectural patterns \cite{iomt_wear2018}.  

\subsection{Healthcare Interoperability and Data Management}
Interoperability in healthcare is critical for enabling seamless data exchange between disparate systems and facilitating integrated patient care \cite{dohertyKeepingPaceWearables2024}. However, integrating wearable data into existing healthcare systems presents unique challenges due to the heterogeneity of data formats, the high frequency of data generation, and the need for real-time processing \cite{mandel2016smart}. Standards such as HL7 Fast Healthcare Interoperability Resources (FHIR)\footnote{\url{https://www.hl7.org/fhir/}} have emerged to provide a common framework for representing and exchanging healthcare data and their alignment with clinical terminology and ontologies \cite{duda2022hl7}. However, existing solutions often focus on specific aspects of interoperability, such as data mapping or API development \cite{junghoon2025design}, or on other data modalities without providing a holistic approach to managing high-volume, heterogeneous data streams from wearables \cite{benhamida2020saecg}. To cover the entire lifecycle of wearable data in clinical settings, frameworks are needed that can harmonize heterogeneous wearable data into standardized formats while maintaining data quality and ensuring compliance with regulations.

\subsection{Wearable Data Processing}
The potential of wearable data in clinical practice is significant, ranging from public health surveillance and prevention to real-time clinical decision support \cite{ming_continuous_2020}. However, there is still a lack of comprehensive frameworks for processing and analyzing such data in a scalable manner.
Existing solutions often focus on specific use cases, such as remote patient monitoring, without providing a generalizable approach to handling the diverse data generated by different wearable devices combined with clinical routine data \cite{ferguson2022effectiveness}.
Wearable data processing is often performed on-device or through vendor-specific platforms, which may not align with clinical data (quality) standards or do not integrate seamlessly with hospital systems to retrieve patient information (e.g., demographics, medical history, or current medications) \cite{shukla2023real, dinh2019wearable}.
Along with wearable data processing, the interplay of patient and case management, a device integration model, and extensibility considerations should be taken into account.
In light of modern cloud and data engineering principles, there is an opportunity to design frameworks that can efficiently ingest, process, and analyze wearable data in real-time, while ensuring interoperability with existing healthcare systems through standardized data models (i.e., FHIR) and supporting machine learning applications for clinical decision support and research.

\section{System Design}
\label{sec:system-design}

\noindent In this section, we provide a high-level overview of the architecture and design principles of our proposed cloud-native framework for wearable data analytics.

\subsection{Overview}
\begin{figure*}[!ht]
    \centering
        \includegraphics[
        width=\linewidth,
        keepaspectratio
    ]{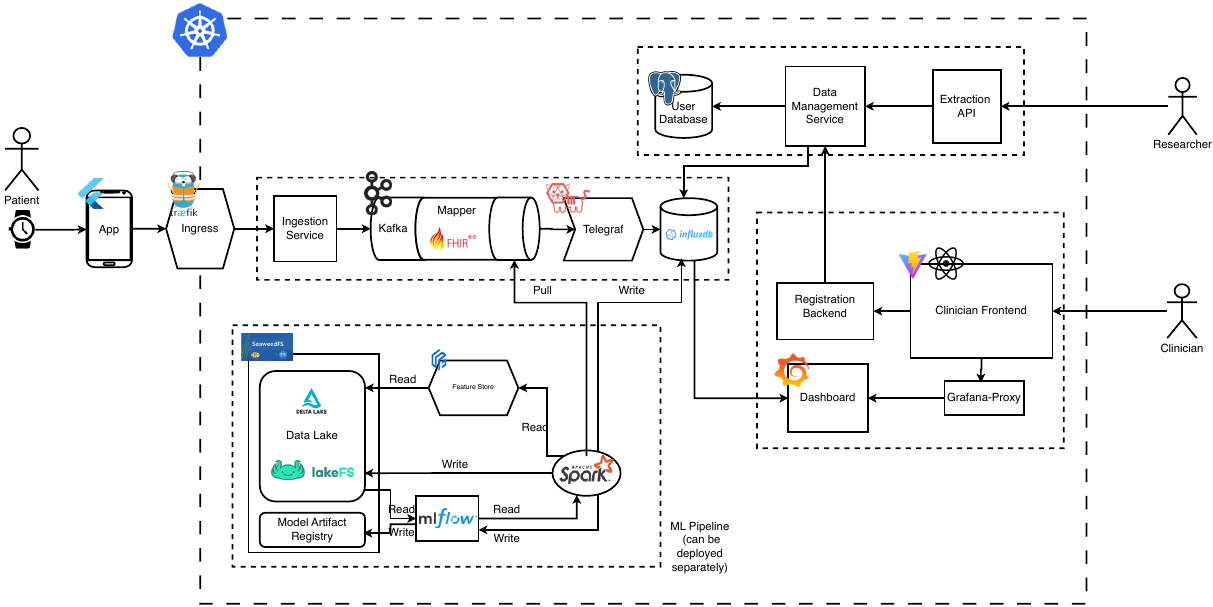}
    \caption{Architecture overview.}
    \label{fig:architecture}
\end{figure*}

To synthesize the requirements and challenges outlined above, we designed a modular, event-driven architecture for clinical workloads. We organized the system into several layers: an acquisition layer for ingesting raw data from wearable devices; a transformation layer for harmonizing heterogeneous data into standardized formats; a storage layer for efficient data management; a presentation layer for clinical visualization and interaction; and an analytics layer for machine learning and research applications. In Figure~\ref{fig:architecture}, we illustrate the high-level architecture of the system. 

At the entry point, the acquisition layer leverages cross-platform mobile applications to collect measurements via native operating system APIs. These data streams are subsequently ingested into a containerized environment for downstream processing. Next, the data transformation layer standardizes and validates these measurements against the FHIR standard, producing unified, interoperable records. Once transformed, the data is stored in specialized components within the data storage and access layer, which separates high-volume time-series wearable data from relational metadata and configuration information. Alongside this pipeline, the platform includes a dedicated data analytics and machine learning layer that enables feature engineering and model development based on measurements collected from a single patient or entire study cohorts. Finally, user-facing services in the presentation layer assemble and visualize the processed data for clinicians, researchers, and developers through dashboards, web interfaces, and queryable APIs.

\subsection{Design principles}
First, we aim for \textbf{modularity} through a microservice infrastructure. Orchestrating containerized components with Kubernetes allows for decoupled development and independent lifecycle management of each architectural layer. This modular approach ensures that the platform remains extensible and observable.

Second, we enable \textbf{scalability} via a distributed event-streaming backbone. High throughput is required to support large-scale hospitals with thousands of concurrent patients. By decoupling high-frequency data ingestion from downstream processing, the system handles variable workloads through horizontal resource scaling. This architecture supports both real-time stream processing for immediate insights and batch processing for complex, longitudinal analysis across cohorts. 

Third, we prioritize \textbf{optimized data persistence} through a tiered storage strategy. High-velocity time-series data is routed through a low-latency hot path for rapid access, while historical records are moved to a data lake for long-term storage. This separation ensures system performance and cost-efficiency as study cohorts and data volumes expand.

Fourth, we propose a privacy-preserving design and prepare for \textbf{regulatory compliance} in hospital settings. The system implements Role-Based Access Control (RBAC) to enforce the principle of least privilege, ensuring that access to sensitive health data is strictly regulated to clinicians, researchers, or developers. Data security is further reinforced through in-transit encryption and logical segregation across all layers.

Fifth, we only use open-source software to enable full \textbf{transparency and data sovereignty}, especially in on-premise secure processing environments.

\subsection{Unified Data Model}

Large-scale clinical workloads require a unified data model to facilitate seamless analysis across diverse cohorts \cite{keoghBreakingDigitalFortress2024}. 
However, consumer wearables are inherently heterogeneous, differing in sensors, sampling windows, and proprietary derivation algorithms \cite{dohertyKeepingPaceWearables2024, liHeartRateVariability2023}. 
Consequently, interoperability poses both a syntactic mapping challenge, i.e., reconciling field names and units,  and a semantic one: determining clinical and technical comparability \cite{jamiesonGuideConsumergradeWearables2025}. 
This is further complicated by inconsistent validation across the consumer market. 
We categorize these challenges into four areas:

\begin{enumerate}
\item \textbf{Metric Heterogeneity:} Wearables utilize diverse sensing modalities (e.g., PPG, ECG, HR, SpO$_2$, accelerometry). Measurement availability varies significantly across manufacturers and device generations \cite{dohertyKeepingPaceWearables2024, fullerReliabilityValidityCommercially2020, keoghBreakingDigitalFortress2024}.
\item \textbf{Variable Accuracy:} Performance depends on the manufacturer, activity type, and user characteristics. While heart rate is generally reliable, metrics like energy expenditure and $\mathrm{VO_2}$ may exhibit higher variability \cite{germiniAccuracyAcceptabilityWristWearable2022, choeAppleWatchAccuracy2025}. Additionally, physiological factors can impair accuracy, such as skin pigmentation for PPG measurements \cite{singhImpactSkinPigmentation2024}.
\item \textbf{Semantic Divergence:} Identical labels often mask different underlying definitions or aggregations. For instance, heart rate variability (HRV) may represent RMSSD or SDNN \cite{liHeartRateVariability2023,jamiesonGuideConsumergradeWearables2025}. Summary measures are typically shaped by proprietary filtering and quality control rather than standardized clinical protocols \cite{fullerReliabilityValidityCommercially2020}.
\item \textbf{Rapid Iteration Cycles:} Hardware and firmware updates frequently alter device behavior, often outpacing peer-reviewed validation. Thus, longitudinal comparability within the same product line is not guaranteed \cite{abimansourAccuracyRoleConsumer2024}. The same concern applies to mobile operating system APIs, whose updates may silently alter data access behaviour or available data types.
\end{enumerate}

To meet the design principles and address the challenges outlined above, we propose a vendor-agnostic system for full data sovereignty.
Rather than developing custom integrations with each wearable device, the proposed system instead relies on the health OS frameworks on iOS and Android phones, as virtually all major wearable manufacturers already maintain integrations with them. 
This approach not only broadens platform support but also reduces the maintenance burden associated with these integrations, which is of particular relevance in a production environment. 
Data from the health frameworks are captured through a custom-built mobile application that runs in the background and periodically extracts raw data, which are then transmitted to the event-driven backend for transformation and storage. 

After ingestion into a stream-processing engine, device-specific measurements are normalized into FHIR to provide a common medical representation across heterogeneous data sources. 
FHIR was selected because it is widely adopted in clinical environments, commonly expressed in \texttt{JSON}, supported by mature parsing and validation libraries, and sufficiently expressive to represent clinically relevant measurements without requiring the system to encode domain-specific medical semantics.
In addition, to ensure clinical validity, the system must preserve comprehensive metadata, including device generation, firmware version, and measurement context, alongside each normalized record.

The transformation pipeline is designed around two objectives: First, the incoming data must be mapped to FHIR through a declarative and readily configurable specification and validated to ensure that invalid data cannot enter the system. 
Second, the resulting FHIR payloads must be handled efficiently so that storage and throughput remain scalable under concurrent load. 
To handle the heterogeneous data emitted by the devices in a continuously running production setting, we introduce a novel declarative approach. 
That is, developers specify how raw data fields correspond to FHIR elements. 
Furthermore, field-level transformations can be applied to unify data from different sources into a common format.
This also enables comprehensive data lineage tracking.

To prevent invalid clinical data from entering the pipeline, FHIR validation is integrated into the transformation stage and performed locally using a standards-compliant validator\footnote{\url{https://build.fhir.org/validation\#validators}}.
Validation is substantially more computationally expensive than transformation, so the system supports a configurable validation sampling rate via an environment variable (see \cref{sec:evaluation}). 

To satisfy the previously identified demand for efficient FHIR data processing, we propose using Kafka Streams to perform the transformation and validation stages.
It is tightly integrated into the Kafka ecosystem and offers high throughput, scalability, and fault tolerance through a simple interface, reducing complexity when working in a cloud-native environment.
Together with Kafka's pub-sub model, this allows messages to be processed flexibly by keeping mapping logic isolated from the transport layer and providing strong horizontal scalability.

Beyond processing, storage efficiency presents a further challenge at scale.
Raw FHIR data in \texttt{JSON} format exhibit significant redundancy and can consume considerable storage at scale. 
To mitigate excessive storage consumption and enable scalability across hundreds of concurrent users, FHIR data are stored in minimized form rather than in a document store, leveraging (i) dependency-tree structures and (ii) columnar storage to reduce storage footprint. 
The dependency tree is automatically inferred from the declarative mapping description (via the \texttt{YAML} file.
It is subsequently used to determine the minimal subset of values required to derive all other fields upon export or retrieval.
As the data requirements may change in continuously running production environments, dependency trees are versioned.
For columnar storage, we use InfluxDB and InfluxDB Line Protocol.
After dependency-based minimization, FHIR data are transformed to InfluxDB Line Protocol via the aforementioned declarative YAML specifications and written to InfluxDB, which enables efficient columnar storage and retrieval.
This data structure is optimized for high write speeds while still providing minimal storage occupation.
During extraction at a later point in time, the reconstruction process traverses the dependencies and persisted mapping-schema versions to efficiently regenerate the full FHIR structure without requiring re-derivation of the dependency tree at runtime.

\section{Implementation}
\label{sec:implementation}

\begin{figure*}
    \centering
    \includegraphics[width=\linewidth]{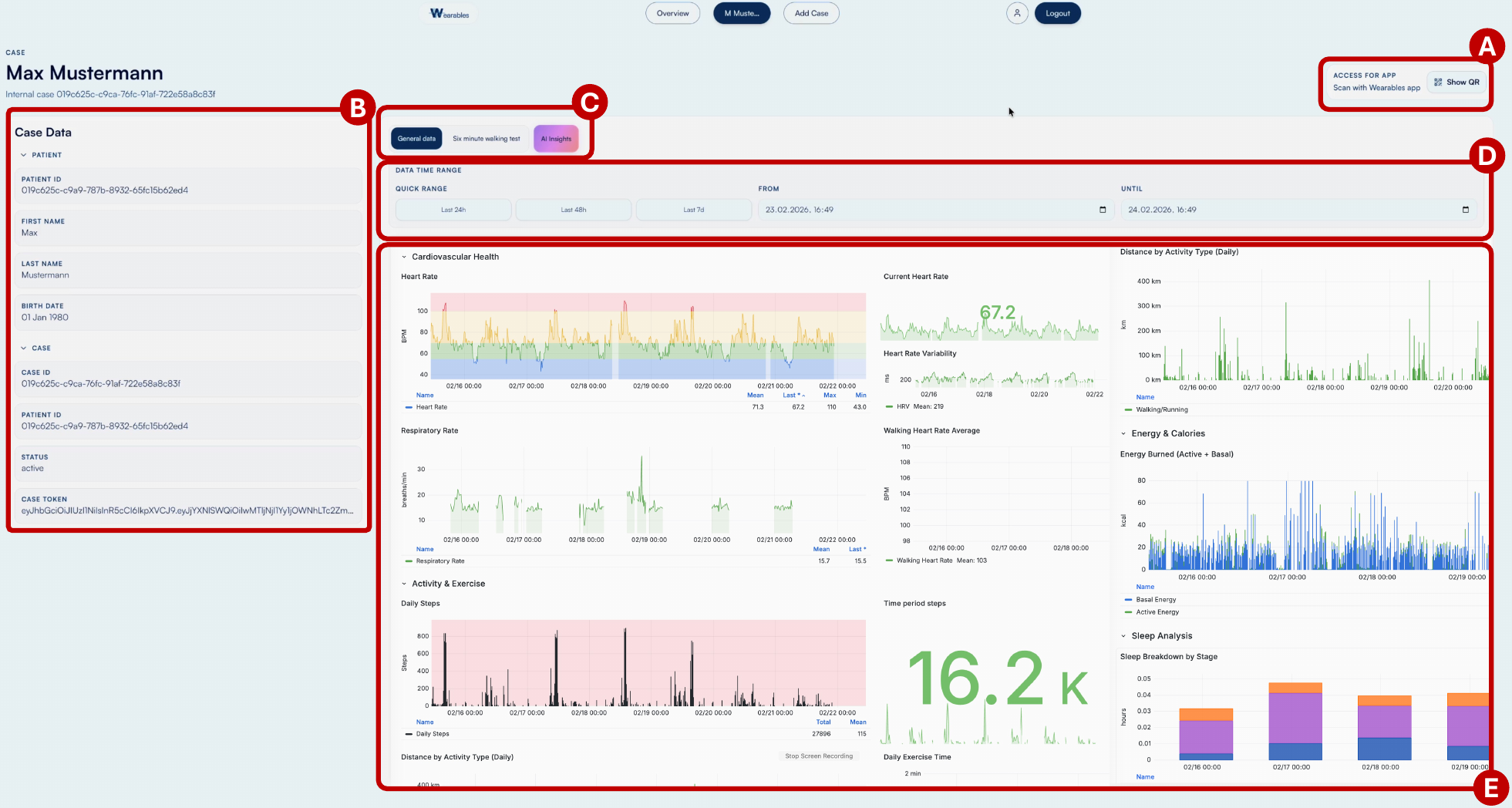}
    \caption{Clinician's view on collected data. 
    \textbf{A}: QR code for connecting a device to the patient case; 
    \textbf{B}: patient demographics; 
    \textbf{C}: data view selector incl. AI analytics; 
    \textbf{D}: time frame selector for the current view; 
    \textbf{E}: data visualizations of the selected view.}
    \label{fig:frontend}
\end{figure*}

\noindent The following sections provide a detailed description of the implementation of the transformation layer, including the storage optimizations and their interplay. 
Further, we discuss the design of a dedicated machine learning pipeline that operates on the transformed data.

\subsection{Data Acquisition \& Ingestion Layer}
We built a cross-platform mobile application with Flutter that serves as the primary data acquisition interface, collecting raw data from wearable devices, packaging them into structured payloads, and transmitting them to the backend ingestion pipeline. 
Rather than communicating directly with wearable hardware over Bluetooth, it leverages the health data aggregation APIs of the host operating system (Apple HealthKit on iOS and Google Health Connect on Android). 

Measurements are categorized into three types: instantaneous (point-in-time readings such as heart rate or body temperature), cumulative (values aggregated over an interval, such as step counts or calories burned), and duration-based (activities spanning a defined time range, such as sleep stages or workout sessions). 
Manual and automatic background synchronization are supported, with the latter running at the maximum frequency permitted by the OS.

The Import Service, implemented in Python using FastAPI and deployed as a containerized service, acts as a thin validation and routing gateway between mobile clients and the internal event streaming infrastructure. 
It fulfills three responsibilities: (i) authenticating requests by verifying a JWT Bearer token, issued via the frontend during case pairing and encoding patient and case identity claims, against its signature, issuer, and expiration; (ii) validating incoming JSON payloads against a defined schema; and (iii) publishing validated payloads to the \texttt{wearables-raw} Kafka topic for consumption by downstream components such as the FHIR Mapper and Validator. 
The Kafka producer is configured with a 50~ms linger time and a 64~KB batch size for throughput optimization, and employs exponential backoff retry logic for transient failures.
It is implemented as a lazily initialized singleton (\texttt{MeasurementProducer}), maintaining a persistent broker connection that is reused across requests to avoid per-request connection overhead and to enable efficient batching and compression.
This asynchronous design decouples ingestion from processing, allowing the service to respond to the client immediately after validation, reducing the utilization of limited background computation time on the client device.

\subsection{Event-driven Transformations}\label{ssec:mapper}

\begin{figure}[ht]
    \centering
    \includegraphics[width=\linewidth]{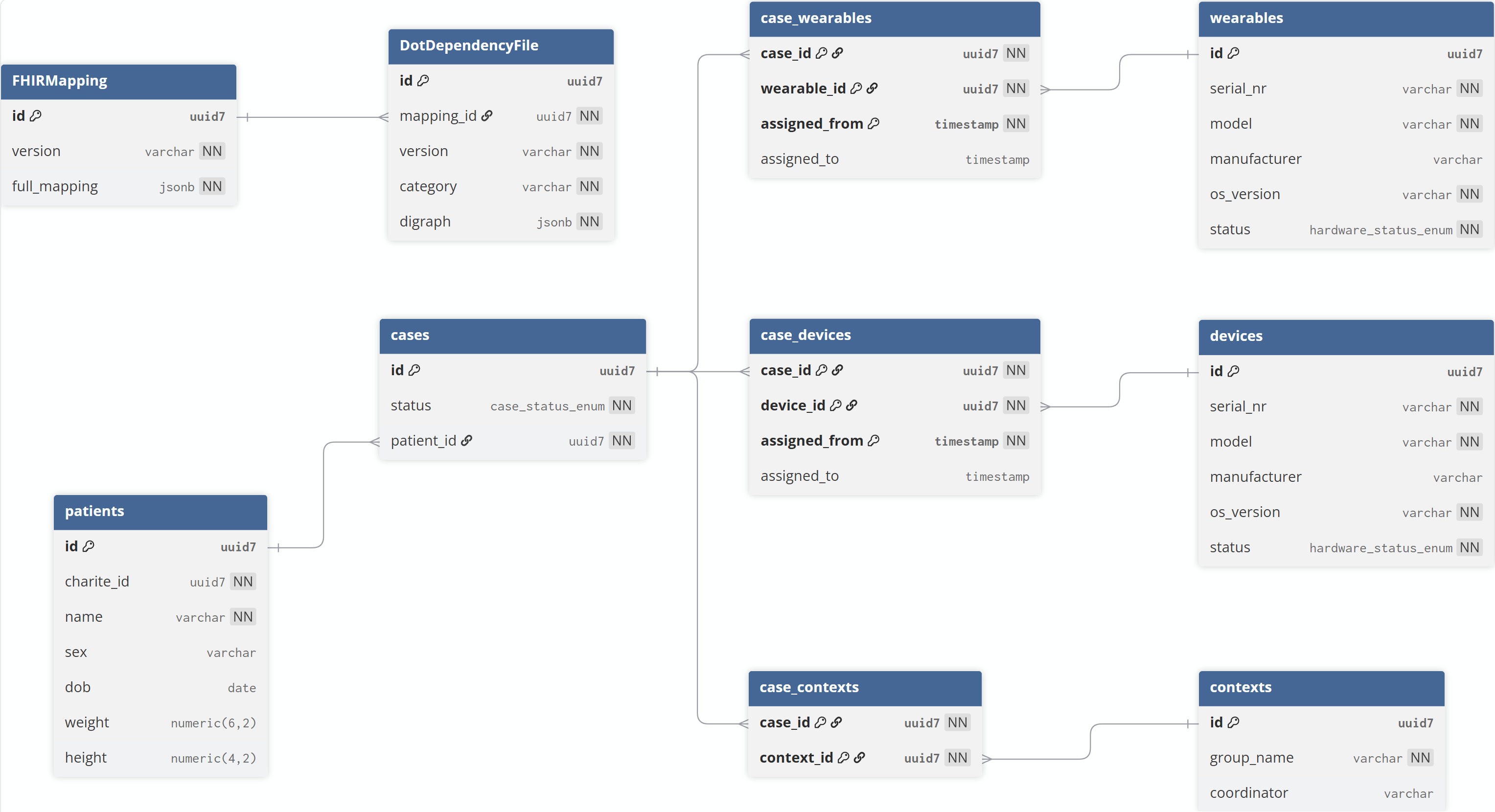}
    \caption{Entity-relationship diagram of the PostgreSQL schema.}
    \label{fig:er-diagram}
\end{figure}

The transformation pipeline is built on Kafka and Kafka Streams.
After ingestion into Kafka, the Kafka Streams instances (hereafter referred to as Mapper) receive these messages.
As these messages are sent to the ingestion service in batches and hence forwarded to Kafka in batches, they are split into individual measurements, re-keyed, and repartitioned to a dedicated topic to improve throughput by parallelizing stateless transformations. 
Because the messages are independent, they can be processed by a horizontally scaled fleet of Mapper instances.
The number of instances and the number of threads that run on each instance are configurable.
However, maximal parallelism is subject to the constraint that the product of Mapper instances and threads per instance does not exceed the number of partitions of the single measurement topic.%
For instance, a system comprising 3 Kafka Streams instances, each running 10 threads, enables 30 requests to be processed concurrently.
Once a measurement is fully processed, it is delivered to the \texttt{wearables-lp} topic.
Telegraf bridges Kafka and InfluxDB, consuming from the \texttt{wearables-lp} topic and writing to the database.

The Mapper fulfills three tasks for each measurement.
First, it maps incoming raw data to FHIR according to a declaratively defined schema.
The mapping is declared via a user-defined YAML file that is loaded into the Mapper at startup. 
From that schema, the system infers a dependency tree among the FHIR fields at startup to optimize storage usage.
Second, the resulting FHIR resource is validated locally using a standards-compliant validator.
Validation is substantially more computationally expensive than transformation, so the system supports configurable sampling via an environment variable. 
In this mode, every \textit{n}-th measurement is validated while a per-instance counter preserves the stateless processing model of the stream processors \cite{abraham_scuba_2013}.
If validation fails, the message is written to a Dead Letter Queue (DLQ), which is monitored or can be investigated manually.
The third step was motivated by the finding that storing FHIR resources in raw form was infeasible due to storage size.
Here, the message is compressed by using the dependency tree to infer the minimal subset of fields from which all other fields can be deduced.
Subsequently, these fields are mapped to the Influx Line Protocol to exploit columnar storage and further reduce storage footprint.
The minimization procedure accounts for the types of dependencies between fields, ensuring that the original FHIR resource can always be fully reconstructed.

The storage layer is split into two databases:
As wearable telemetry arrives as high-throughput, irregularly sampled time-series data, InfluxDB was chosen as a storage solution because it is designed specifically for timestamped data with efficient support for time-windowed filtering and retrieval.

Alongside the time-series path, PostgreSQL serves as the system's relational store for the platform's structured relational and configuration data, comprising core domain entities, temporal assignment records, patient and case structures, hardware inventory, reusable grouping information, and versioned mapping artifacts referenced during FHIR reconstruction.

Data from both databases are accessed through a dedicated data management service.
The core domain entities are illustrated in the entity-relationship diagram shown in~\cref{fig:er-diagram}.

\subsection{Data Presentation \& Dashboards}

The web application serves as the central access point to the platform, following a component-based architecture in which the accompanying backend service acts as an integration boundary between the frontend and the underlying data sources. In \cref{fig:frontend}, we show a single patient's overview.
The backend service is implemented as a Node.js/TypeScript service using the Express framework, with its API specified through an OpenAPI contract, and manages the full authentication lifecycle, including login, registration, magic-link verification, and logout.
Access to interface components is governed by the roles assigned to a user account.
Clinicians are provided with an overview of all patient cases imported into the platform, with support for search and sort to quickly locate a specific case; cases not yet available can be imported directly from a connected Hospital Information System.
The visualisation layer is built around Grafana, which is embedded directly into the web application to provide a unified interface for clinicians. Grafana serves as the primary tool for rendering time-series health data from wearable devices and was chosen for its mature, configurable dashboard environment and native InfluxDB integration.

\subsection{Analytics \& Machine Learning Workloads}\label{ssec:ml-pipeline}
Building an effective analytics platform requires a foundation of high-quality, well-structured data, which is particularly challenging for wearable devices that generate high-frequency signals that are often noisy, incomplete, or biased.
To address this challenge, we adapt the medallion architecture and, to the best of our knowledge, present the first real-time implementation of this pattern for wearable data.
The pipeline is implemented using Spark Structured Streaming, which treats a live data stream as an unbounded table to which new rows are continuously appended, decomposing queries into incremental micro-batch executions that are identical in form to batch queries.

The pipeline is organized into three layers.
In the Bronze layer, a dedicated Spark streaming job \cite{spark_streaming} reads messages from the \texttt{wearables-lp} Kafka topic and writes them to SeaweedFS (via its S3-compatible interface), preserving raw messages while appending only basic ingestion metadata.
In the Silver layer, a second job transforms the raw Bronze data into a structured Delta Lake table stored in lakeFS on a dedicated branch, extracting measurement names, device tags, metric values, and timestamps into a consistent schema.
In the Gold layer, a third job connects the Silver data to the machine learning feature pipeline: it filters relevant measurements, converts each micro-batch into feature sets, and pushes them to a Feast online store via its push API.
The job then immediately retrieves the latest features from Feast to ensure that inference uses exactly the same feature definitions and transformations applied during training, thereby preventing training-serving drift.

We propose embedding a production ML model directly into the Gold-layer job, loaded from a model registry (in our implementation, MLflow) using a versioned alias.
Designing new prediction models falls outside the scope of this paper; however, for demonstration purposes, we implemented a simplistic \texttt{models:/heart\_rate\_model@Production} model to demonstrate the capabilities of our architecture.
Assuming heart rate prediction were a clinically relevant task, the job would perform real-time inference and write predictions together with their metadata to the \texttt{ml-predictions} InfluxDB bucket as time-series points, making them directly accessible in the Grafana dashboards described above.

Model versioning is managed through the alias mechanism in MLflow.
By assigning the \texttt{Production} alias to a registered model, both training jobs and serving infrastructure share a unified reference.
Downstream consumers automatically resolve the updated alias at runtime when a new model is promoted, requiring no configuration changes.
The Gold-layer job maintains its own checkpoint in a dedicated lakeFS branch, thereby isolating its internal state from the rest of the pipeline.

The metadata backend of SeaweedFS relies on PostgreSQL, which introduces a relational bottleneck for metadata-intensive workloads involving directories containing millions of small files.
While a distributed metadata engine such as TiKV is recommended for production environments exceeding 10\,TB, the current PostgreSQL configuration remains effective for workloads below this threshold.

A reliable machine learning system in healthcare demands rigorous monitoring, versioning, and reproducibility across datasets and models.
In particular, continuous monitoring for data drift remains an important avenue for future research.

\subsection{Interoperability \& Extensibility}
The platform adopts FHIR as its canonical data exchange format and exposes well-defined APIs, ensuring compatibility with established healthcare infrastructure such as Hospital Information Systems and patient data management platforms, while providing a foundation for extending support to additional wearable devices, measurement types, or clinical workflows.
The data management service exposes a REST interface with resource-based CRUD endpoints, supporting filtering, sorting, and cursor pagination on list routes.
Workflow endpoints around the case model allow patients to be queried with their cases, and cases to be retrieved in an expanded form embedding related entities such as the linked patient, contexts, or hardware assignments.
Telemetry is accessible via structured endpoints, which group InfluxDB rows into logical items, and raw endpoints, which expose individual field rows directly; both support time-range bounds, measurement, field, and tag filters, and window pagination.
Alongside REST, a GraphQL API enables flexible read access across related entities following the Relay node and connection model~\cite{hartig2018semantics}, with dedicated telemetry query and subscription fields supporting both InfluxDB tag filters and PostgreSQL-based entity filters.

A dedicated extraction service is decoupled from the data management service to enforce a clear boundary between clinical and research data access paths, exposing three access patterns: a paginated JSON endpoint with cursor-based navigation filterable by measurement type, time range, patient, device, and case; a streaming CSV export that iterates pages internally and streams results row by row, suitable for exports spanning millions of records; and a FHIR export endpoint that transforms internal measurements into FHIR Observation resources returned as a FHIR Bundle.

\subsection{Regulatory Compliance \& Data Sovereignty}
To comply with foundational privacy principles and provisions of the General Data Protection Regulation (GDPR), the platform implements advanced privacy measures. 
Role-based access control is implemented at the API level. Clinicians and researchers can only access data assigned to their roles. Further access control, e.g., based on study purposes or attribute-level (e.g., specific vital signs), can be easily extended.
We can further apply the principle of data minimization to the APIs. For instance, through our GraphQL implementation, we can massively reduce the given access to wearable data on an attribute level. As opposed to traditional FHIR applications \cite{mandel2016smart}, this greatly reduces over-fetching.
Our architectural design also aligns with cloud-native privacy engineering principles \cite{grunewald2021cloud}. Among others, we added a CI/CD pipeline for continuous security scans on the used third-party libraries.
Furthermore, data sovereignty is maintained by ensuring that the entire stack is deployable within a private cloud environment, thus preventing clinical data from traversing across non-trusted infrastructure.

\subsection{Code Availability}

\faGithub~\textbf{We release our development as open source software,}%
\footnote{\url{https://github.com/WearableAnalytics}}
being dedicated to open source and open science principles, and fostering collaboration on wearable data processing.

\begin{figure*}[!h]
    \centering

    \begin{minipage}[t]{0.32\textwidth}
        \centering
        \vspace{0pt}
        (a)\\[2pt]
        \includegraphics[height=3.2cm,keepaspectratio]{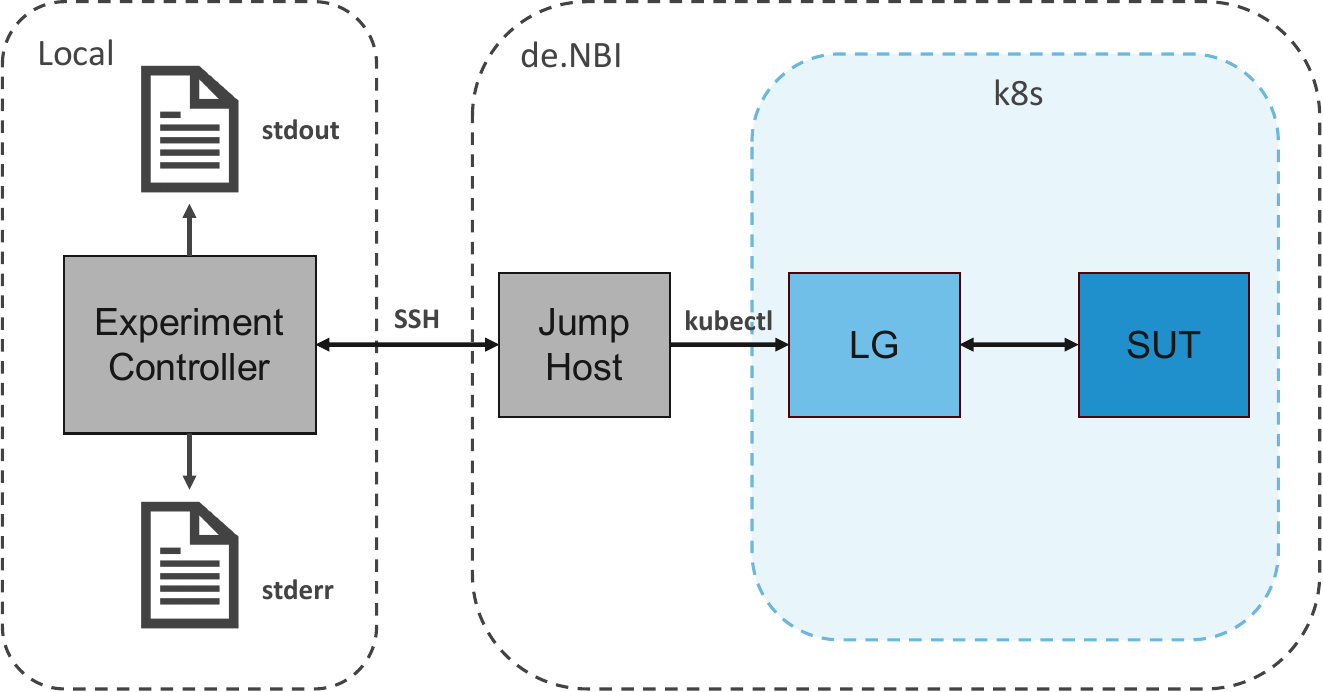}
    \end{minipage}
    \hfill
    \begin{minipage}[t]{0.32\textwidth}
        \centering
        \vspace{0pt}
        (b)\\[2pt]
        \includegraphics[height=3.2cm,keepaspectratio]{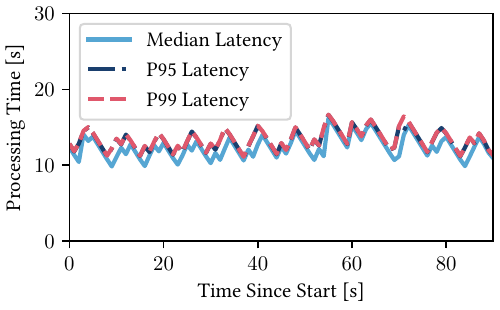}
    \end{minipage}
    \hfill
    \begin{minipage}[t]{0.32\textwidth}
        \centering
        \vspace{0pt}
        (c)\\[2pt]
        \includegraphics[height=3.2cm,keepaspectratio]{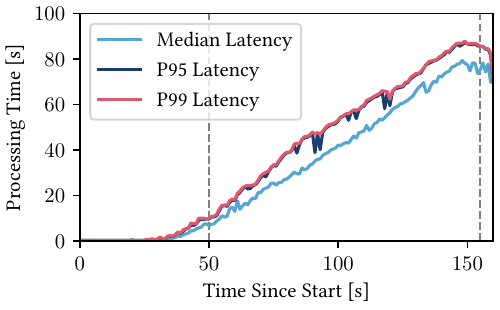}
    \end{minipage}

    \par\vspace{0.8em}\par

    \begin{minipage}[t]{0.48\textwidth}
        \centering
        \vspace{0pt}
        (d)\\[2pt]
        \includegraphics[height=3.2cm,keepaspectratio]{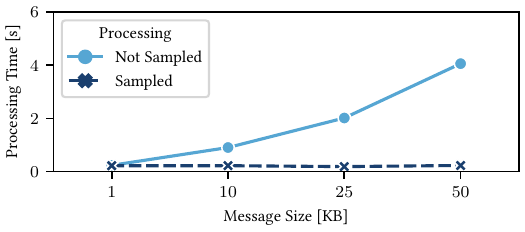}
    \end{minipage}
    \hfill
    \begin{minipage}[t]{0.48\textwidth}
        \centering
        \vspace{0pt}
        (e)\\[2pt]
        \includegraphics[height=3.2cm,keepaspectratio]{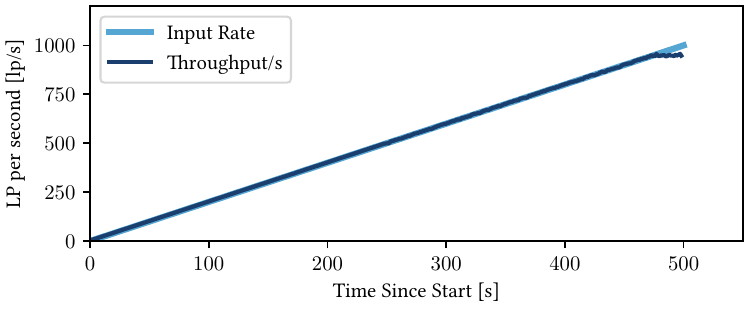}
    \end{minipage}

    \caption{
    \textbf{(a)} Experimental setup. 
    \textbf{(b)} Processing time for a single line-protocol inference at approximately 100 messages per second (msg/s) during the 90 s experiment phase. 
    \textbf{(c)} 50th, 95th, and 99th latency percentiles during a three-phase experiment: 50 s ramp-up to 50 req/s, 100 s constant load at 50 req/s, and 5 s ramp-down to 0 req/s.
    \textbf{(d)} Processing time for different validation approaches. 
    \textbf{(e)} Request rate vs. throughput, ramping up to 1000 line-protocol requests per second over 500 s.
    }
    \label{fig:system_eval}
\end{figure*}

\section{Evaluation}
\label{sec:evaluation}

In the following section, we evaluate whether the prototype can handle realistic clinical workloads. %

\subsection{Experimental Setup \& Goal}

In line with state-of-the-art benchmarking practice~\cite{bermbach_cloud_2017}, our setup (see \cref{fig:system_eval}a) consists of four core components: the System Under Test (SUT), a Load Generator, an Experiment Controller, and a measurement storage component. The SUT comprises three \textit{de.NBI-medium}\footnote{\label{note:flavors}\url{https://cloud.denbi.de/wiki/Concept/flavors/}} virtual nodes, each equipped with an AMD~EPYC~7002 (Rome-series) processor with 16~CPU cores and 16~GB of RAM, deployed within the German academic OpenStack environment \textit{de.NBI}.
The Load Generator runs on a dedicated benchmark node provisioned on an Intel Xeon (Ice Lake) processor with 8~CPU cores and 8~GB of RAM.
All nodes are deployed within the same region (\textit{de.NBI Berlin}), with system clocks synchronized via a local NTP server. Each experiment is managed by a controller
responsible for the full experiment life cycle.

Our evaluation focuses on two main performance aspects of the platform, that have a direct impact on user experience and clinical feasibility:
First, we evaluate performance of the ML inference by measuring processing time per inference and throughput in relation to input rate.
Importantly, our evaluation does not test the predictive performance of the machine learning models. 
Instead, it focuses on the performance of the underlying infrastructure. 
Second, we evaluate the performance of the hot ingestion path of the system. 
Here, we measure end-to-end latency (i.e. time from ingestion into Kafka until successful write to InfluxDB) and the impact of sampling rate on processing time.

\subsection{Machine Learning Performance Results}

As shown in \cref{fig:system_eval}b processing time remains stable throughout the experiment, ranging between 10 and 15 seconds per inference.
The oscillation of latency observed is a direct consequence of Spark's Structured Streaming micro-batch processing model. 
Due to the usage of Spark's \texttt{default triggers}, the pipeline processes data in batches while running as soon as data arrives. 
As described in \cref{ssec:ml-pipeline}, the pipeline consists of three stages, each with approximately 3 to 5 seconds of processing time. 
This results in a general processing latency of 9 to 15 seconds per inference.%

\cref{fig:system_eval}e visualizes throughput under a linearly increasing input rate that reaches up to 1000 line-protocol messages per second. 
We find that the pipeline scales linearly with the input rate to approximately 980 messages per second. 
This exceeds our requirements and demonstrates the capability of the pipeline to reliably serve larger workloads. 
This is enabled through Apache Spark's concurrent architecture, which effectively handles high load by distributing computation across available resources \cite{spark}.

\subsection{Hot Path Performance Results}

To evaluate hot path performance, we evaluate latency in a stress-test scenario with a load of 50 messages per second, each containing 10\,KB of measurement data.
This corresponds to approximately 100 data points, yielding a steady load of 5000 measurements per second, an unusually high workload for a system of this scale.
As shown in \cref{fig:system_eval}c, the pipeline processes up to 30 messages per second without a noticeable performance drop. 
Beyond this threshold, processing times strongly increase, reaching a maximum latency of 85\,s per message.
Two primary bottlenecks are identified: (i) the Kafka Streams Mapper, whose FHIR validation introduces severe processing overhead, and (ii) Telegraf, which is configured to write batches of 1000 Line Protocol messages per tick (100\,ms), becoming a limiting factor under high load; both warrant further investigation.

Moreover, FHIR validation (see \cref{ssec:mapper}) dominates processing time, as shown in \cref{fig:system_eval}d.
Applying validation to every message leads to increased latency with larger message sizes, causing backpressure and message buildup in Kafka.
In contrast, sampling-based validation significantly reduces this overhead, resulting in nearly constant processing times across message sizes, thereby stabilizing throughput and preventing backlog formation.
Overall, the Mapper achieves an average processing time of approximately 200\,ms per message, while allowing flexible tuning of validation intensity based on system requirements.

\section{Discussion \& Conclusion}
\label{sec:discussion}

Our microservice-based architecture represents a critical shift in handling high-velocity physiological data. By utilizing a distributed event-streaming, our framework effectively decouples the \enquote{hot path} of real-time ingestion from the \enquote{cold path} of long-term research analytics. This separation ensures that the performance of real-time clinical monitoring is not degraded by concurrent, resource-intensive batch processing for longitudinal studies. Furthermore, the ability to horizontally scale the transformation layer (incl. FHIR mapping) allows the system to absorb the bursty nature of wearable data synchronization \cite{ferguson2022effectiveness}, in which multiple devices may offload hours of buffered data simultaneously, without introducing significant latency into the clinical dashboard.

A primary contribution of this work is the novel dependency-aware FHIR minimization scheme, which addresses a long-standing tension between medical interoperability and storage scalability. While FHIR is the gold standard for clinical data exchange, its verbose JSON structure is inherently inefficient for high-frequency time-series data, often leading to a 10–20x increase in storage requirements compared to raw binary formats. Our approach mitigates this by persisting only the unique, non-derivable data points within a time-series data store. By reconstructing the full FHIR resource on-demand via versioned declarative mappings, the framework achieves the best of both worlds, i.e., a low-footprint storage backend optimized for rapid query performance and a standards-compliant interface for external hospital information systems.

Altogether, the platform demonstrates a viable foundation for real-time wearable data collection and clinical integration, yet several directions remain for future work. For instance, device coverage should be expanded beyond platforms with native Apple HealthKit or Google Health Connect integration to include consumer wearables lacking such support, as well as medical-grade devices capturing continuous digital biomarkers such as continuous glucose monitoring.
Advanced privacy measures should be implemented to further partition data across clinical studies or departments. In addition, data minimization techniques, for instance, noise injection or selective field stripping, would further enhance study workflows. The framework enables the development of clinical decision support systems (CDSS) that can utilize raw, high-resolution data that is typically discarded or summarized by existing approaches.
To improve scalability, additional performance benchmarks are required, particularly for real-world machine learning workloads, and the platform would benefit from a comprehensive observability stack. 

Technical integration alone, however, is insufficient to realize clinical value.
The platform should first be leveraged as a dedicated research infrastructure, enabling prospective clinical studies and controlled data collection under defined protocols, including user studies with relevant stakeholders \cite{izmailova2018wearable}.
The ultimate goal is deployment in routine clinical practice, which requires participatory workflow redesign involving clinicians, patients, and technical staff to define how the system is operated, maintained, and iteratively improved.
Once embedded in care delivery, the accumulating real-world data provides a rich substrate for observational research and hypothesis generation at scale, while preserving full data sovereignty. Eventually, these are essential steps towards the Internet of Medical Things (IoMT), enabling data-driven clinical decision-making \cite{iomt_wear2018}.

In summary, we propose an event-driven, cloud-native architecture for collecting, harmonizing, and presenting wearable health data for clinical workflows.

\section*{Acknowledgments}
This work is partly funded by the \enquote{Wearables: From pilot phase to patient care} project at Charité -- Universitätsmedizin Berlin.

This work was supported by the de.NBI Cloud within the German Network for Bioinformatics Infrastructure (de.NBI) and ELIXIR-DE (Forschungszentrum Jülich and W-de.NBI-001, W-de.NBI-004, W-de.NBI-008, W-de.NBI-010, W-de.NBI-013, W-de.NBI-014, W-de.NBI-016, W-de.NBI-022).

GPT-5 was used to improve the grammar and spelling of the manuscript. The AI system was also used to review the text for clarity and coherence; however, all content was generated by the authors. Coding contributions were partially assisted by GitHub Copilot and are documented in the corresponding commit messages or code comments. The authors take full responsibility for the content of this article, including any errors or inaccuracies.

\bibliographystyle{IEEEtran}
\bibliography{bibliography}

\end{document}

%% file: header.tex
\documentclass[conference]{IEEEtran}
\IEEEoverridecommandlockouts
\usepackage{cite}
\usepackage{amsmath,amssymb,amsfonts}
\usepackage{algorithmic}
\usepackage{graphicx}
\usepackage{textcomp}
\usepackage{csquotes}
\usepackage{comment}
\usepackage{xcolor}
\usepackage{subcaption}
\usepackage{todonotes}
\usepackage[T1]{fontenc}
\usepackage[scaled=0.95]{inconsolata}

\usepackage{fontawesome5}

\usepackage[hyphens]{url}

\usepackage{xcolor}
\usepackage[colorlinks=true,
            linkcolor=blue,
            citecolor=blue,
            urlcolor=blue]{hyperref}
\usepackage[capitalise,nameinlink]{cleveref}

\crefname{figure}{Figure}{Figures}

  \usepackage{tikz}

\def\BibTeX{{\rm B\kern-.05em{\sc i\kern-.025em b}\kern-.08em
    T\kern-.1667em\lower.7ex\hbox{E}\kern-.125emX}}

\usepackage{pbalance}